\documentclass[aps,prb,twocolumn,showpacs,amsmath,amssymb]{revtex4}
\usepackage{epsfig}
\usepackage{graphicx}
\usepackage{booktabs}
\usepackage{multirow}
\usepackage{tabularx}
\usepackage{times}
\usepackage{color}

\definecolor{Red}{rgb}{1,0,0}

\definecolor{Blu}{rgb}{0,0,01}

\definecolor{Green}{rgb}{0,1,0}

\newcommand{\be}{\begin{equation}}
\newcommand{\ee}{\end{equation}}

\DeclareMathOperator{\sgn}{sgn}

\begin{document}

\title{Josephson effect in S/F/S junctions: spin bandwidth asymmetry vs. Stoner exchange}

\author{Gaetano Annunziata,$^{1}$ Henrik Enoksen,$^{2}$ Jacob Linder,$^{2}$ Mario Cuoco,$^{1}$ Canio Noce,$^{1}$ Asle Sudb{\o}$^{2}$}

\affiliation{$^{1}$ SPIN-CNR, I-84084 Fisciano (Salerno), Italy \\
Dipartimento di Fisica ``E. R. Caianiello'', Universit\`a di
Salerno, I-84084 Fisciano (Salerno), Italy}

\affiliation{$^{2}$ Department of Physics, Norwegian University of
Science and Technology, N-7491 Trondheim, Norway}


\begin{abstract}
We analyze the dc Josephson effect in a ballistic
superconductor/ferromagnet/superconductor junction by means of the
Bogoliubov–-de Gennes equations in the quasiclassical Andreev
approximation. We consider the possibility of ferromagnetism
originating from a mass renormalization of carriers of opposite
spin, i.e.~a spin bandwidth asymmetry. We provide a general
formula for Andreev levels which is valid for arbitrary interface
transparency, exchange interaction, and bandwidth asymmetry and
analyze the current-phase relation, free energy, and critical
current, in the short junction regime. We compare the phase
diagrams and the critical current magnitudes of two identical
junctions differing only in the mechanism by which the mid-layer
becomes magnetic. We show that a larger number of $0-\pi$
transitions caused by a change in junction width or polarization
magnitude is expected when ferromagnetism is driven by spin
bandwidth asymmetry compared to Stoner magnetism. Moreover, we
show that these features can be present also for ferromagnets of
the Stoner type having only a partial bandwidth asymmetry.
\end{abstract}

\date{\today}

\maketitle

\section{INTRODUCTION}
The antagonistic nature of ferromagnetism and singlet
superconductivity notwithstanding, coexistence of these phenomena
is not entirely ruled out and their interplay may in fact yield
quite rich physics. Coexistence of ferromagnetism and
superconductivity, predicted to occur in a phase where both the
gauge and the continuous translational symmetries are
spontaneously broken simultaneously,~\cite{Fulde,Larkin} is
suggested to emerge in many theoretical
models.~\cite{Maska,Romano2010} However, a decisive experimental
verification of such coexistence is still lacking. On the other
hand, the interplay between ferromagnetism and superconductivity
in heterostructures has been the foundation for developments of
numerous novel experimental techniques over the last decade or
so.~\cite{Soulen98,Tedrow94} Besides the interest from a
fundamental physics viewpoint, such systems hold potential in
terms of future spintronics~\cite{spintronicsreview} applications.
The versatility of heterostructures involving superconductors and
ferromagnets stems from the basic fact that Andreev
reflections,\cite{Andreev64,Beenakker92} processes taking place at
the interfaces converting quasiparticles into Cooper pairs, are
strongly affected by spin polarization.~\cite{deJong,
linder_prb_07}

In this paper, we focus on ballistic
superconductor/ferromagnet/superconductor (S/F/S) Josephson
junctions with conventional superconducting leads. These are
systems capable of sustaining a supercurrent carried by Cooper
pairs in the superconducting leads and by quasiparticles in the
ferromagnetic mid-layer. The unique interplay between ferromagnetic
and superconducting orders provides quasiparticles with extra
phase shifts absent in junctions with a non-magnetic mid-layer.
This can give rise to the appearance of a so-called
``$\pi-$phase''.~\cite{Bulaevskii} Under such circumstances, the
energy minimum of the junction is reached at a phase difference of
$\pi$ across the junction, unlike the standard ``$0-$phase'' in
junctions with a non-magnetic mid-layer. The existence of the
$\pi-$phase has been experimentally
confirmed.~\cite{Ryazanov,Kontos} As far as potential applications
are concerned, $\pi$-junctions are considered to be candidates for
realizing solid state qubits.~\cite{Yamashita}

Previous theoretical studies on S/F/S
junctions~\cite{Chtchelkatchev,Radovic,Cayssol,Petkovic,linder_prl_08,Millis88,Fogel2000,Barash2002,josreview,CPRreview}
have described the F layer using the Stoner model of metallic
ferromagnetism, with oppositely polarized carriers occupying
rigidly shifted bands. However, the interplay of Coulomb repulsion
and the Pauli principle driving a metal into a ferromagnetic
state,~\cite{Abinitio} may induce a {\it spin-dependent
renormalization of the masses of charge carriers} with opposite
spins, i.e. a spin bandwidth asymmetry. This mechanism appears in
microscopic approaches where off-diagonal terms of Coulomb
repulsion, generally neglected in studies based on the Hubbard
model, are taken into account. A mean-field treatment of these
contributions in addition to the exchange and nearest-neighbor
pair hopping terms shows that quasiparticle energies for the two
spin species are not simply splitted but get different bandwidths,
i.e.~effective masses. The net effect of these interactions is to
render the hopping integral in the kinetic term spin dependent
through bond charge Coulomb repulsion terms which are different
for spin up and down carriers.~\cite{Hirsch} For low enough
temperature and depending on Hamiltonian parameters, ferromagnetic
order can be established only through this spin bandwidth
asymmetric renormalization.~\cite{Hirsch99} In this picture the
ferromagnetism should be understood as kinetically driven, in the
sense that it arises from a gain in kinetic energy rather than
potential energy, unlike the usual Stoner scheme. We notice that
this mass-split metallic ferromagnetism has been experimentally
found to be the origin of the optical properties of the colossal
magnetoresistance manganites,~\cite{Okimoto} of some rare-earth
hexaborides~\cite{hexab} as well as of some magnetic
semiconductors.~\cite{Singley} Spin bandwidth asymmetry will
substantially affect the coexistence of ferromagnetism and
superconductivity,~\cite{Ying} as well as the proximity
effect~\cite{Cuoco} and transport in F/S
bilayers.~\cite{Annunziata,Annunziata2010} It is also responsible
for an extension of the regime in which FFLO (Fulde-Ferrell-Larkin-Ovchinnikov) phase can be
stabilized in heavy-fermion systems.~\cite{Kac2009,Maska2010} In
this work, we analyze the consequences of this unconventional
magnetism in ballistic S/F/S Josephson junctions by solving the
Bogoliubov-de Gennes equations~\cite{BdG} for arbitrary spin
polarization and interface transparency.

We show that bandwidth asymmetry in the F layer of a S/F/S
junction modifies many physical properties such as Andreev levels
dispersion, Josephson current and free energy. We provide a
general formula for Andreev levels which holds for arbitrary
interface transparency, exchange interaction, and bandwidth
asymmetry and we calculate the corresponding Josephson current in
the short junction regime for some important limiting cases.
Comparing the results for a spin bandwidth asymmetry ferromagnet
with those obtained for a conventional Stoner ferromagnet, we
demonstrate analytically that the former manifests features that
are only quantitatively different from the latter for low degrees
of polarization. We show numerically that these differences become
qualitative in the intermediate/high polarization regime. Then,
evaluating the phase diagrams we infer that the mere mechanism by
itself is able to shift the ground state superconducting phase
difference from 0 to $\pi$ or viceversa, with a spin bandwidth
asymmetry ferromagnet tending to increase the number of possible
$0-\pi$ transitions in a fixed range of parameters.

This paper is organized as follows. In Section II, we introduce
the model and notation, and provide a general formula for Andreev
levels in S/F/S which is valid for arbitrary interface
transparency, exchange interaction, and bandwidth asymmetry. In
Section III we present an analytical form for Andreev levels and
Josephson current in different limiting cases and numerically
analyze the same quantities, critical current, and free energy for
different polarization, temperature, F layer width, and interface
transparency values. This analysis is performed considering both
Stoner and spin bandwidth asymmetry magnetism in the F layer. The
relationship between the spin polarization and the magnetic
microscopic parameters is clarified by means of a general formula
holding for a generic ferromagnet in which both mechanisms are
present. In this ``mixed'' case, we also show that the features
distinguishing pure spin bandwidth asymmetry from pure Stoner
ferromagnets may exist even though the spin bandwidth asymmetry
contributes only partially to spin polarization. We present phase
diagrams and analyze $0-\pi$ transitions which are driven by
variations in the polarization and width of the F layer. The
conclusions and a discussion on the applicability of our results
to real materials is presented in Section IV, while analytical
details for some quantities used in the paper are reported in the
Appendices.

\section{MODEL AND FORMALISM}

\begin{figure}
\begin{center}
\includegraphics[width=0.4\textwidth]{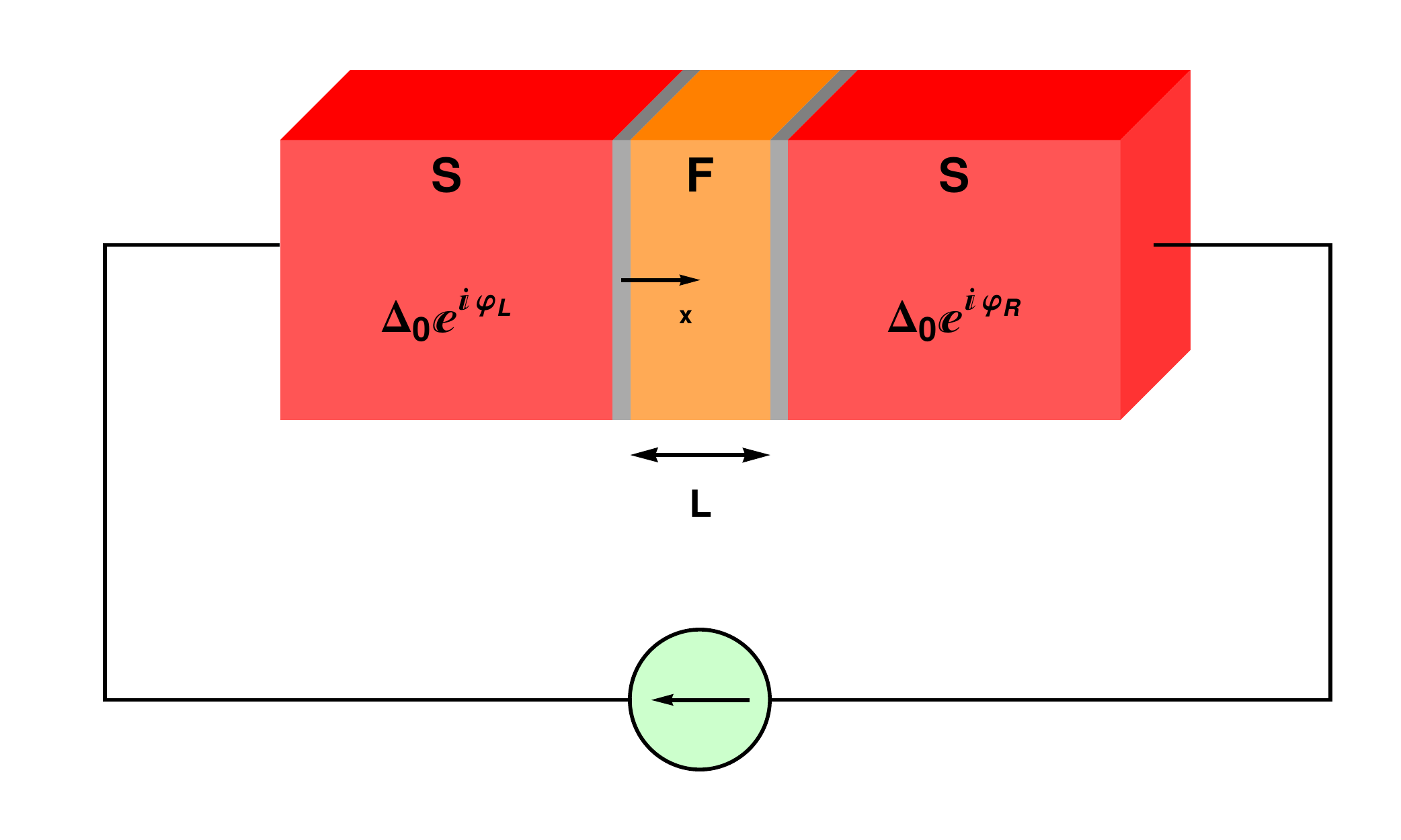}
\caption{A sketch of the current-biased S/F/S junction
analyzed in the paper, along with notations that are used. The
superconductors are treated as reservoirs, whereas the magnetic
mid-layer has a finite width $L$.} \label{fig:sketch}
\end{center}
\end{figure}

We consider a ballistic Josephson junction composed of a
ferromagnetic layer of width $L$ sandwiched between two
conventional $s-$wave singlet superconducting electrodes (see
Fig.~\ref{fig:sketch}). Since the S electrodes are isotropic, we
will consider an effective one dimensional model. Moreover, our
results are independent on the direction of the magnetic moment
direction in F. The propagation of quasiparticles is described by
the Bogoliubov-de Gennes equations~\cite{BdG}
\be
\left( \begin{matrix} H_0^\sigma && \rho_{\sigma} \Delta \\
\rho_{\sigma} \Delta^\ast && -H_0^{\bar{\sigma}} \end{matrix}
\right) \left(
\begin{matrix} u_\sigma \\ v_{\bar{\sigma}} \end{matrix} \right)
=\varepsilon \left( \begin{matrix} u_\sigma \\ v_{\bar{\sigma}}
\end{matrix} \right), \,\sigma=\uparrow,\downarrow\quad,
\label{eq:Bdg} \ee
where $\bar{\sigma}=-\sigma$,
$\rho_{\uparrow(\downarrow)}=+1(-1)$, and
$\left(u_\sigma,v_{\bar{\sigma}}\right)\equiv\psi_\sigma$ is the
energy eigenstate in the electron-hole space associated with the
eigenvalue $\varepsilon$. The single-particle Hamiltonian is \be
H_0^\sigma= H_L+H_F^\sigma+H_R+H_I,  \ee where
\begin{eqnarray}
H_L  &=&  \left[ -\hbar^2 \mathbf{\nabla}^2/2m-E_F \right]
\Theta(-x)  \nonumber \\
H_F^\sigma & = & \left[
-\hbar^2\mathbf{\nabla}^2/2m_\sigma-\rho_\sigma
U-E_F\right]\Theta(x) \Theta(L-x)  \nonumber \\
H_R  &=& \left[ -\hbar^2 \mathbf{\nabla}^2/2m-E_F
\right]\Theta(x-L) \nonumber \\
H_I  &=& H \left[ \delta(x)+\delta(x-L) \right]. \label{eq:Ham}
\end{eqnarray}
Here, different effective masses $m_\sigma$ for $\sigma$-polarized
particles in the F layer, mimicking bandwidth asymmetry, has been
included and $U$ is the exchange interaction, $E_F$ is the Fermi energy,
$\Theta(x)$ is the Heaviside step function, $m$ is the effective
mass of the quasiparticles in the superconductors, and the
parameter $H$ quantifies scattering strength at S/F and F/S
interfaces. We assume rigid superconducting order parameters with
equal gap amplitude on both sides of the junction, i.e.~$\Delta=\Delta_0 \left[ e^{i \varphi_L} \Theta(-x)+ e^{i
\varphi_R} \Theta(x-L) \right]$, where $\varphi_{L(R)}$ is the
phase of the left(right) superconductor. When considering finite
temperature properties, we will assume the usual BCS dependence
and let $\Delta_0\rightarrow\Delta_0 \tanh \left(1.74 \sqrt{T_c /T
-1} \right)$, where $T_c$ is the superconducting critical
temperature.

Employing the quasiclassical Andreev
approximation,~\cite{Andreev64} the solution of
Eqs.~\eqref{eq:Bdg} may be written as

\begin{widetext}
\[
\psi_\sigma (x) =
\begin{cases}
a_\sigma \left(\begin{matrix} \rho_{\sigma} v \\ u \end{matrix} \right) e^{ikx} +  b_\sigma \left(\begin{matrix} u \\ \rho_{\sigma} v \end{matrix} \right) e^{-ikx},& x < 0 \\
\left[ \alpha_\sigma  e^{i q_\sigma x}+
\beta_\sigma  e^{-i q_\sigma (x-L)} \right] \left(\begin{matrix} 1\\0 \end{matrix} \right) +
\left[  \gamma_\sigma  e^{i q_{\bar{\sigma}} (x-L)}+
             \delta_\sigma  e^{-i q_{\bar{\sigma}} x}   \right]  \left(\begin{matrix} 0\\1 \end{matrix} \right)    ,&  0<x<L \\
    c_\sigma \left(\begin{matrix} u \ e^{i \varphi/2}\\ \rho_{\sigma} v\  e^{-i \varphi/2}\end{matrix} \right) e^{ik(x-L)} +  d_\sigma \left(\begin{matrix} \rho_{\sigma} v\  e^{i \varphi/2}\\u \ e^{-i \varphi/2}\end{matrix} \right)e^{-ik(x-L)},&  x > L
    \end{cases}
\]
\end{widetext}
where
\begin{align}
u = \sqrt{\frac{\varepsilon+\sqrt{\varepsilon^2-\Delta_0^2}}{2\varepsilon}},\; v =
\sqrt{\frac{\varepsilon-\sqrt{\varepsilon^2-\Delta_0^2}}{2\varepsilon}}
\; ,
\end{align}
and $k=\sqrt{2 m E_F}/\hbar$, $q_\sigma=\sqrt{2 m_\sigma
(E_F+\rho_\sigma U)}/\hbar$ are Fermi wavevectors in the S and F
electrodes respectively, and $\varphi = \varphi_R-\varphi_L$. The
coefficients $a_\sigma$, $b_\sigma$, $c_\sigma$, and $d_\sigma$
are the probability amplitudes for Andreev reflection as a hole-like quasiparticle
(HLQ), normal reflection as an electron-like quasiparticle (ELQ), transmission to the
right electrode as an ELQ, and transmission to the right electrode
as an HLQ respectively, while the coefficients $\alpha_\sigma$, $\beta_\sigma$,
$\gamma_\sigma$, and $\delta_\sigma$ are associated with right- and
left-going ELQ and HLQ propagating through the ferromagnetic
layer. All these probability amplitudes can be calculated by
solving the linear system resulting from the boundary conditions
\begin{eqnarray}
\psi_\sigma(0_-) & = & \psi_\sigma(0_+) \nonumber \\
\psi_\sigma(L_-) & = & \psi_\sigma(L_+) \nonumber \\
- \frac{2H}{\hbar^2}\psi_\sigma(0) &=&
\frac{1}{m}\psi_\sigma'(0_-)-\left(\begin{matrix}
1/m_\sigma\\1/m_{\bar{\sigma}}
\end{matrix} \right)
\psi_\sigma'(0_+)   \nonumber \\
- \frac{2H}{\hbar^2}\psi_\sigma(L) &=& \left(\begin{matrix}
1/m_\sigma\\1/m_{\bar{\sigma}}
\end{matrix} \right)\psi_\sigma'(L_-)-\frac{1}{m}
\psi_\sigma'(L_+), \label{eq:boundaries}
\end{eqnarray}
which are valid for a general ferromagnetic electrode displaying
both exchange splitting and bandwidth asymmetry. When the latter
mechanism is present, electron- and hole-like parts of the wave
function derivatives have to be joined separately with different
masses. This is analogous to the situation where the insulating
barriers are polarized with magnetic moment parallel to the one in
the F layer and thus act with different strength on particles with
opposite spin.~\cite{AnnunziataSUST} When only exchange splitting
is present, i.e.~$m_\sigma=m_{\bar{\sigma}}=m$, the usual form of
the boundary conditions is recovered.~\cite{Radovic}

Inserting the wave function into the boundary conditions, we
obtain a homogenous system of linear equations for the scattering
coefficients. By imposing that the system has non-trivial
solutions, we can obtain a relation between energy and phase
difference such that the coherent subgap processes can effectively
take place, i.e.~Andreev levels with dispersion
$\varepsilon_i(\varphi)$, $i=\{1,...,4\}$.~\cite{Beenakker912} We
are able to provide a general form of the levels which holds for
arbitrary transparency of the insulating barriers, polarization in
the ferromagnetic layer, and relative weight of the exchange and
bandwidth asymmetry contributions. We find
$\varepsilon_i(\varphi)=
\varepsilon_\sigma^\pm(\varphi)=\pm\varepsilon_\sigma(\varphi)$
and
\begin{widetext}
\be
 \varepsilon_{\sigma}(\varphi) =
 \Delta_0\sqrt{\frac{A^2+B^2-A\left(C+D\cos^2(\varphi/2)\right)
  +\rho_{\sigma}\sqrt{B^2\left(A^2 + B^2 - \left(C + D\cos^2(\varphi/2)\right)^2\right)} }
  {2\left(A^2+B^2\right)}}, \label{eq:levels}
  \ee \end{widetext}
where $A,B,C,D$ depend on all junction parameters. Their explicit
form will be given in the Appendix~\ref{app1}. Here, we merely
note that for transparent interfaces, and in the absence of
exchange field and bandwidth asymmetry, Eq.~\eqref{eq:levels}
reduces to the well known result $\varepsilon=\Delta_0 \cos
(\varphi /2)$.~\cite{Kulik}

We will focus on the short junction regime, i.e.~$L\ll\xi$ where
$\xi$ is the superconducting coherence length, such that the
Josephson current can be estimated by only considering subgap
contributions by Andreev levels~\cite{Beenakker912,Beenakker91} as
\begin{eqnarray}
I(\varphi) & = & \frac{2 e}{\hbar} \sum_i f(\varepsilon_i)
\frac{d \,
\varepsilon_i }{d \varphi} \nonumber \\
& = & -\frac{2 e}{\hbar} \sum_\sigma \tanh \left( \frac{\beta
\varepsilon_\sigma}{2 } \right) \frac{d \, \varepsilon_\sigma }{d
\varphi}, \label{eq:current}
\end{eqnarray}
where $f(x)$ is the Fermi-Dirac distribution function and  $\beta
=1/k_BT$.

Finally,  the critical current is defined as $I_c={\rm{max}}|I(\varphi)|$ and the
phase of the junction ($0$ or $\pi$) is determined by the minimum of the $\varphi$
dependent part of the free energy
\begin{eqnarray}
F(\varphi) & = & - \frac{1}{\beta} \ln \left[ \prod_i \left(
1+e^{- \beta  \varepsilon_i (\varphi) } \right) \right] \nonumber \\
& = & - \frac{1}{\beta} \sum_\sigma \ln \left[ 2 \cosh \left(
 \frac{\beta \varepsilon_\sigma (\varphi) }{2 } \right) \right].
\label{eq:freeenergy}
\end{eqnarray}

\section{RESULTS}
In this section we analyze Andreev levels, current-phase relation,
free energy, and critical current for a short S/F/S junction. Our
model enables us to take into account both bandwidth asymmetry and
exchange interaction in the F layer. Defining the polarization as
$M=(n_\uparrow - n_\downarrow)/(n_\uparrow + n_\downarrow)$ and
integrating densities of states as calculated from the $H_F^\sigma$
term in Eq.~\eqref{eq:Ham}, we find that in one dimension and at
$T=0$:
\begin{align}\label{eq:mag}
M=p_+/p_- \text{ with } p_\sigma = -\rho_{\sigma}\Big[1-\rho_{\sigma}\sqrt{\frac{m_\uparrow}{m_\downarrow}
\frac{1+U/E_F}{1-U/E_F}}\Big].
\end{align}
For a fixed value of the exchange splitting $U$, the effect of
mass mismatch is to enhance the net polarization for $m_\uparrow >
m_\downarrow$, and to hinder it the other way around.
Eq.~\eqref{eq:mag} describes a general F where both exchange and
bandwidth asymmetry are present. We notice that Eq.\eqref{eq:mag} reduces to the case where only exchange interaction is present when $m_\uparrow/m_\downarrow=1$ and $U\neq 0$ or the case where only bandwith asymmetry is present when $m_\uparrow/m_\downarrow\neq 1$ and $U=0$. We will refer to these
two cases as a Stoner ferromagnet (STF) and a spin bandwidth asymmetry ferromagnet (SBAF), respectively. When both mechanisms are
present there are several combinations of exchange energy and
mass ratio that give the same polarization $M$. We choose to
describe this mixed case through a parameter $W$ $\in [0,1]$ which
quantifies the relative weight of the two mechanisms without
changing the total polarization in such a way that a pure STF and
a pure SBAF will correspond to $W=0$ and $W=1$, respectively. The
detailed procedure chosen to calculate the value assumed by
exchange and mass mismatch for a given polarization $M$ and degree
of mixture $W$ is illustrated in Appendix~\ref{app2}. In Table~\ref{table}, the magnetic parameter values for a STF, a
SBAF, and a mixture of the two with $W=0.5$, are reported for
three values of the polarization $M=\{0.25, 0.50, 0.75 \}$.
Different Josephson effect features are expected in S/STF/S and
S/SBAF/S junctions, even for an equally polarized F layer. This is
so, since specifying a degree of polarization is equivalent to fix
the ratio of Fermi wavevectors $q_\uparrow / q_\downarrow$.
However, in a SBAF and in a STF the wavevectors are different,
i.e.~$q_\sigma ^{STF} \neq q_\sigma ^{SBAF}$, and accordingly the
center of mass momentum acquired by Cooper pairs will be
different.

\begin{center}
\begin{table}
\begin{tabular}{p{0.1cm} c p{2.5cm} c p{0.5cm} c c c p{0.2cm}}
\hline\hline
 & M                        & &     &&$U/E_F$  & & $m_\uparrow /m_\downarrow$    &     \\
\hline
 &\multirow{3}{*}{0.25}     & & STF  &  &  8/17   & &             1                 &     \\
 &                          & & SBAF  &  &  0      & &          25/9                 &     \\
  &                          & &STF $\&$ SBAF  &  &  0.193      & &          1.878                 &     \\[3ex]
 &\multirow{3}{*}{0.50}     & & STF    &&  4/5    & &             1                 &     \\
 &                          & & SBAF    &&  0      & &             9                 &     \\
  &                          & & STF $\&$ SBAF    &&  0.287      & &             4.988                 &     \\[2ex]
 &\multirow{3}{*}{0.75}     & & STF    &&  24/25  & &             1                 &     \\
 &                          & & SBAF    &&  0      & &             49                &     \\
  &                          & & STF $\&$ SBAF   &&  0.324      & &             24.99                &     \\
\hline\hline
\end{tabular}
\caption{Values of the normalized exchange interaction and mass
asymmetry used for three different polarization values. Here STF
$\&$ SBAF represents a mixture of the two magnetic mechanism with
$W=0.5$.} \label{table}
\end{table}
\end{center}


In the following, we will also consider finite temperature
properties in the range $0\leq T\leq T_c$. Even if
Eq.~\eqref{eq:mag} was derived for $T=0$, it can be used for $T
\leq T_c$, since the polarization varies only slowly on the scale
of the superconducting critical temperature as long as $T_M \gg
T_c$, with $T_M$ being the Curie temperature. This condition is
easily fulfilled in a typical experimental situation where the
Josephson junction is realized by a ferromagnetic transition
metals compound sandwiched between conventional superconductors,
e.g.~Niobium. In our analysis, we fix $\Delta_0=1$ meV and $E_F=5$
eV, consistent with the Andreev approximation. The width of the
ferromagnetic layer $L$ is not fixed. However the maximum value
used in our analysis is such that $L/\xi \approx 0.01$, such that
we are in the short junction regime, corresponding to a width in
the range 1--10 nm. We introduce a dimensionless parameter
quantifying the scattering strength at insulating interfaces as
$Z= 2 m H / \left( \hbar^2 k \right)$.

Before exploring the general case, we look closer at the low
polarization limit for the Andreev levels to get an idea of what
to expect. Expressing the mass mismatch in a SBAF and the exchange
energy in a STF as a function of polarization yields
\begin{align}
    \sqrt{\frac{m_{\uparrow}}{m_{\downarrow}}} &= \frac{1+M}{1-M} \label{eq:y}, \\
    \frac{U}{E_F} &= \frac{2M}{1+M^2} \label{eq:U}.
\end{align}
At first glance, Eqs.~\eqref{eq:y} and~\eqref{eq:U} indicate that
exchange in a STF only depends on odd powers of the polarization
$M$, while mass mismatch in a SBAF depends on all powers.
Furthermore, the expansion of Eq.~\eqref{eq:y} is strictly
positive, while Eq.~\eqref{eq:U} has alternating signs. Due to the
intricate dependence on mass mismatch and exchange energy of the
general Andreev levels Eq.~\eqref{eq:levels} (see the Appendix for
details), both SBAF mass mismatch and STF exchange will depend on
all powers of $M$. However, the different sign behavior may
indicate that SBAF and STF behave differently for large degrees of
polarization. From now on to lighten our notation we will refer to
S/SBAF/S and S/STF/S as simply SBAF and STF. The first evidence of
different behavior is encountered in the $M \ll 1$, $Z \ll 1$
limit. Expanding Eq.~\eqref{eq:levels} around $U/E_F \ll 1$ for
STF and $ \sqrt{m_{\uparrow}/m_{\downarrow}} \approx 1$ for SBAF, we obtain
\begin{align}
    \varepsilon^{\text{STF}}_{\sigma} &= \Delta_0 \left( \cos \left(\frac{\varphi}{2}\right)
    + \rho_{\sigma}\frac{kL}{2}\frac{U}{E_F} \sin\left(\frac{\varphi}{2}\right)\right) 
\label{eq:STFenergy}
\end{align}
for STF and
\begin{align}
    \varepsilon^{\text{SBAF}}_{\sigma} &= \Delta_0 \left( \cos\left(\frac{\varphi}{2}\right) +
    \rho_{\sigma}\frac{kL}{2}\left(\sqrt{\frac{m_{\uparrow}}{m_{\downarrow}}} - 1\right) \sin\left(\frac{\varphi}{2}\right)
    \right)
\label{eq:SBAFenergy}
\end{align}
for SBAF when $Z=0$. Then, for low polarization and including the
first term from the $Z\ll 1$-expansion, both expressions reduce to
the same form
\begin{equation}
    \varepsilon_{\sigma} = \Delta_0 \left(\cos \left( \frac{\varphi}{2} \right) +
    \rho_{\sigma}M \sin \left( \frac{\varphi}{2} \right) \left(kL+\theta Z\sin^{2}kL\right)\right) ,
\label{eq:lowM}
\end{equation}
where $\theta = +1 (-1) $ for STF (SBAF). Thus in the low polarization regime we expect
SBAF and STF to show the same behavior for $Z=0$, while the
$\theta$-factor indicates that they can be different for nearly
transparent interfaces. This is a consequence of a different
interplay between the two ferromagnetic mechanisms and the
insulating barrier strength described by
Eq.~\eqref{eq:boundaries}. By differentiating Eq.~\eqref{eq:lowM}
with respect to $\varphi$ and inserting it into
Eq.~\eqref{eq:current}, we obtain the current-phase relation
\begin{widetext}
\begin{align}\label{eq:CPRlowM}
  \frac{I(\varphi)}{I_0} = 2 \frac{s_1\sin(\varphi/2)\sinh[\beta\Delta_0\cos(\varphi/2)] -s_2 kLM \cos(\varphi/2)\sinh[\beta\Delta_0kLM\sin(\varphi/2)]}{\cosh(\beta\varepsilon_{\uparrow}/2)\cosh(\beta\varepsilon_{\downarrow}/2)} ,
\end{align}
\end{widetext}
where $s_1 = \sgn\left(\cos\left(\varphi/2\right)\right)$, $s_2 =
\sgn\left(\sin\left(\varphi/2\right)\right)$, and $I_0 =
e\Delta_{0}/\hbar$. We have again put $Z=0$ for simplicity.

From Eq.~\eqref{eq:CPRlowM}, we see that the current-phase
relation contains a cosine term in addition to the standard sine
term. One might think that the cosine term will create a
non-vanishing current at $\varphi = 0, \pi$, but this is not the
case. The cosine term will cancel out at $\varphi = 0, \pi$ due to
the Fermi-Dirac distribution function, as shown in
Eq.~\eqref{eq:CPRlowM}. However, in a non-equilibrium situation
where levels can be differently populated, these terms may give
rise to exotic features such as fractional AC Josephson
effect~\cite{kwon,baselmans} or a
$\varphi$-junction.~\cite{goldobin} This possibility is matter of
current investigation.

In the tunneling limit it is possible to obtain an analytical form
for the current for an arbitrary degree of polarization. For $T=0$
and including only the first term in the $Z\gg 1$ expansion, we
find
\begin{equation}
   \frac{I(\varphi)}{I_0} = \frac{1-M^2}{1+M^2} \frac{2 \sin (\varphi) }
   {Z^4 \sin \left(k L  \sqrt{\frac{(1+M)^2}{1+M^2}}   \right)   \sin \left(k L  \sqrt{\frac{(1-M)^2}{1+M^2}}   \right)   }
\label{eq:Istf}
\end{equation}
for STF and
\begin{equation}
    \frac{I(\varphi)}{I_0} = \frac{2 \sin (\varphi) }
   {Z^4 \sin \left(k L  \sqrt{\frac{1+M}{1-M}}   \right)   \sin \left(k L  \sqrt{\frac{1-M}{1+M}}   \right)   }
\label{eq:Isbaf}
\end{equation}
for SBAF. At first glance, one may think that the main difference
between the two cases is the term $(1-M^2)/(1+M^2)$ which is
present only for STF and should damp the current for high degrees
of polarization. We will later see that such an effect exists for
low $Z$. However, for high $Z$ this is masked by the fact that the
current is  small. In fact, the main difference between
Eqs.~\eqref{eq:Istf} and~\eqref{eq:Isbaf} is that the denominator
becomes wildly oscillating for large $M$ for SBAF, but not for STF.
We will see later that this effect is clearly manifested also beyond the
tunneling limit.

\begin{figure}
\begin{center}
\includegraphics[width=0.5\textwidth]{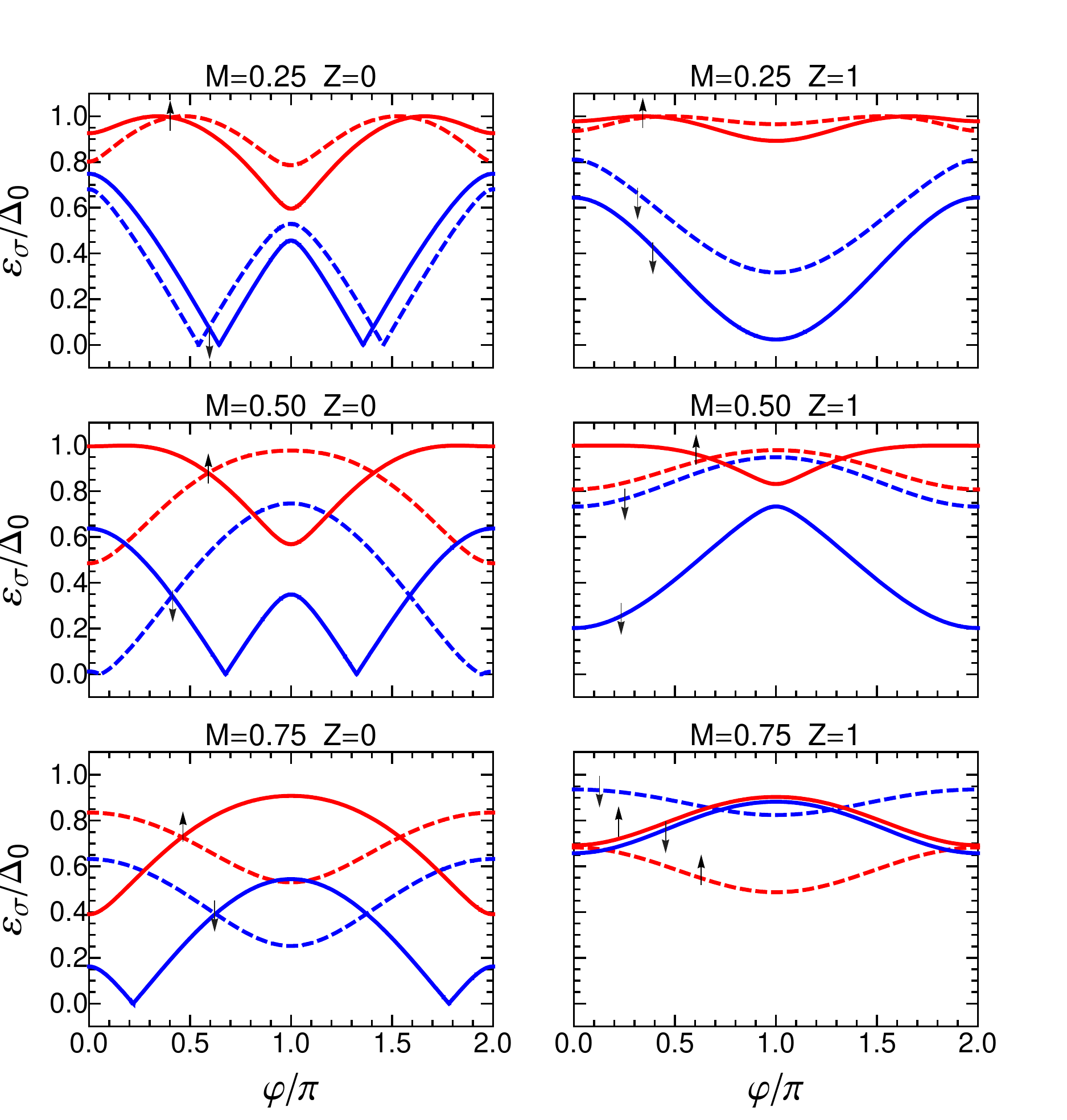}
\caption{(Color online). Positive branches of spin-split
dispersion of the Andreev levels for different values of
polarization and interface barrier strength and for $L k =10$,
$E_F =5$ eV, $\Delta_0 = 1$ meV. Solid lines show levels in SBAF
and dashed lines in STF.} \label{fig:levels}
\end{center}
\end{figure}

We now move on to describe the S/F/S junction properties upon
leaving aside the limits of low polarization and high/low interface
transparency. Fig.~\ref{fig:levels} shows the
$\varphi$ dependence of $\varepsilon_\uparrow$ and
$\varepsilon_\downarrow$ in SBAF (solid lines) and STF (dashed
lines) for transparent (left panels) and insulating (right panels)
interfaces, for three values of polarization $M$. All levels are
flat at $\varphi=0,\pi$, which corresponds to the absence of
Josephson current for these phase difference values. This is a
well known property of S/N/S and S/STF/S junctions, which persists
in the S/SBAF/S case. However, the dispersion of the
$\varepsilon_\sigma$ levels are different for STF and SBAF. This
difference is only quantitative for low polarizations, e.g.~$M=0.25$, as expected from Eq.~\eqref{eq:lowM}. However, the
levels differ qualitatively for higher polarization in terms of their slope and curvature. The order relation between $\varepsilon_\uparrow$ and
$\varepsilon_\downarrow$ can also be different for the two magnetic mechanisms in the F layer, e.g.~for $M=0.75$ and
$Z=1$ $\varepsilon_\uparrow < \varepsilon_\downarrow$ for STF
while $\varepsilon_\uparrow > \varepsilon_\downarrow$ for SBAF.

\begin{figure}
\begin{center}
\includegraphics[width=0.5\textwidth]{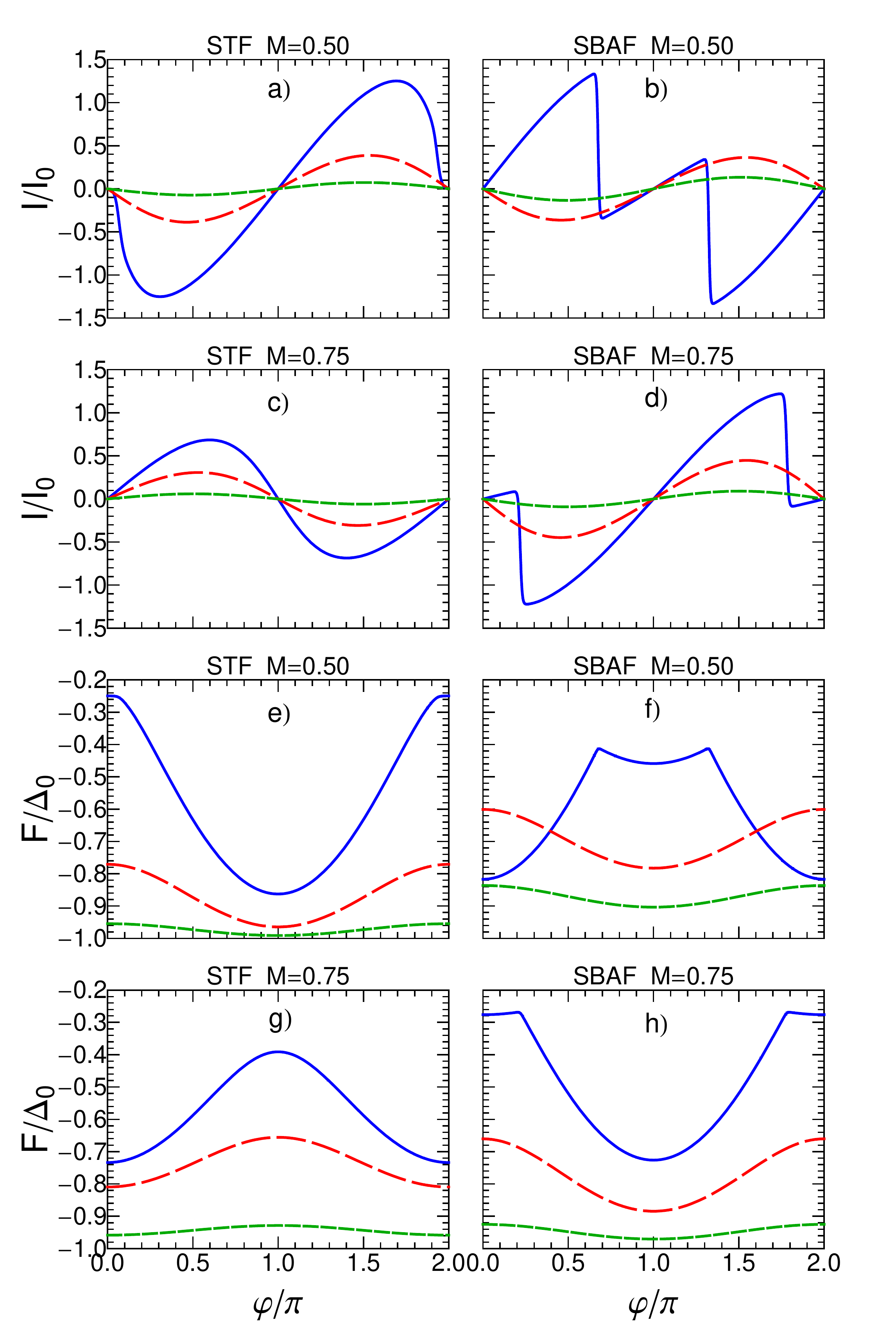}
\caption{(Color online). Current-phase relation and free energy
for STF (left panels) and SBAF (right panels) for intermediate and
high polarization values. In each panel three different interface
barrier strenghts are considered: $Z=0$ (solid lines), $Z=1$
(long-dashed lines) $Z=2$ (short-dashed lines). Here $L k =10$,
$E_F =5$ eV, $\Delta_0 = 1$ meV, $T/T_c=0.01$ are fixed.}
\label{fig:currentenergy}
\end{center}
\end{figure}

These differences affect the Josephson current and free energy, as
shown in Fig.~\ref{fig:currentenergy}, where,  $I/I_0$ and
$F/\Delta_0$ are plotted for different polarization values,
different magnetic mechanisms, and $Z=0$ (solid lines), $Z=1$
(long-dashed lines), $Z=2$ (short-dashed lines). Here $L k =10$,
$E_F =5$ eV, $\Delta_0 = 1$ meV, and $T/T_c=0.01$. For low
polarization values (not reported in the figure) the current and
the energy differ only quantitatively for STF and SBAF, while for
intermediate/high polarization values a qualitative difference
between the effect of the two mechanisms emerges. For instance, at
$M=0.75$ the Josephson current exhibits  a maximum at $\varphi <
\pi$ for STF and $\varphi > \pi$ for SBAF (see panels $c)$ and
$d)$). This means (as shown in panels $g)$ and $h)$) that the energy minimum of the ground-state for the two types of junctions can be
realized at different superconducting phase differences even though
the polarization in the F layer is the same. Moreover, we see that
STF is in a $0-$phase while SBAF is in a $\pi-$phase, so that the
magnetic mechanism by itself is able to drive $0-\pi$ transitions.
In the following, we will refer to the situations where SBAF and
STF are both in $0-$ or $\pi-$phase as ``in phase'', while when
one is in the $0-$phase and the other is in the $\pi-$phase as
``out of phase'' junctions. By inspection of panel $a)$, $b)$,
$e)$, $f)$ of Fig.~\ref{fig:currentenergy} it is also possible to
infer that the interface transparency can play a role in this
bandwidth asymmetry driven phase shift: the STF and SBAF junctions
are out of phase for $Z=0$ and become in phase for $Z=1,2$, since
SBAF undergoes a $0-\pi$ transition for some $Z$ in $[0,1]$. We
underline that STF can also undergo such transitions driven by a
variation in the strength of the barrier, depending on $L$ and $T$
values.

In order to ascertain if the realization of out of phase junctions
is a rarity or a common situation, we have compared the free energy in
SBAF and STF for different width, temperature, degree of spin
polarization, and interface transparency values. Our analysis
shows that the condition of out of phase junctions is mainly
determined by spin polarization: in the low polarization
regime,~i.e. $M\lesssim 0.25$, SBAF and STF junctions very often
share the same phase, while in the intermediate/high polarization
regime,~i.e. $M\gtrsim 0.50$, they are likely to be out of phase. Other
parameters only weakly affect these conditions except interface
transparency which favors out of phase situation whenever the tunneling limit is considered.

\begin{figure}
\begin{center}
\includegraphics[width=0.5\textwidth]{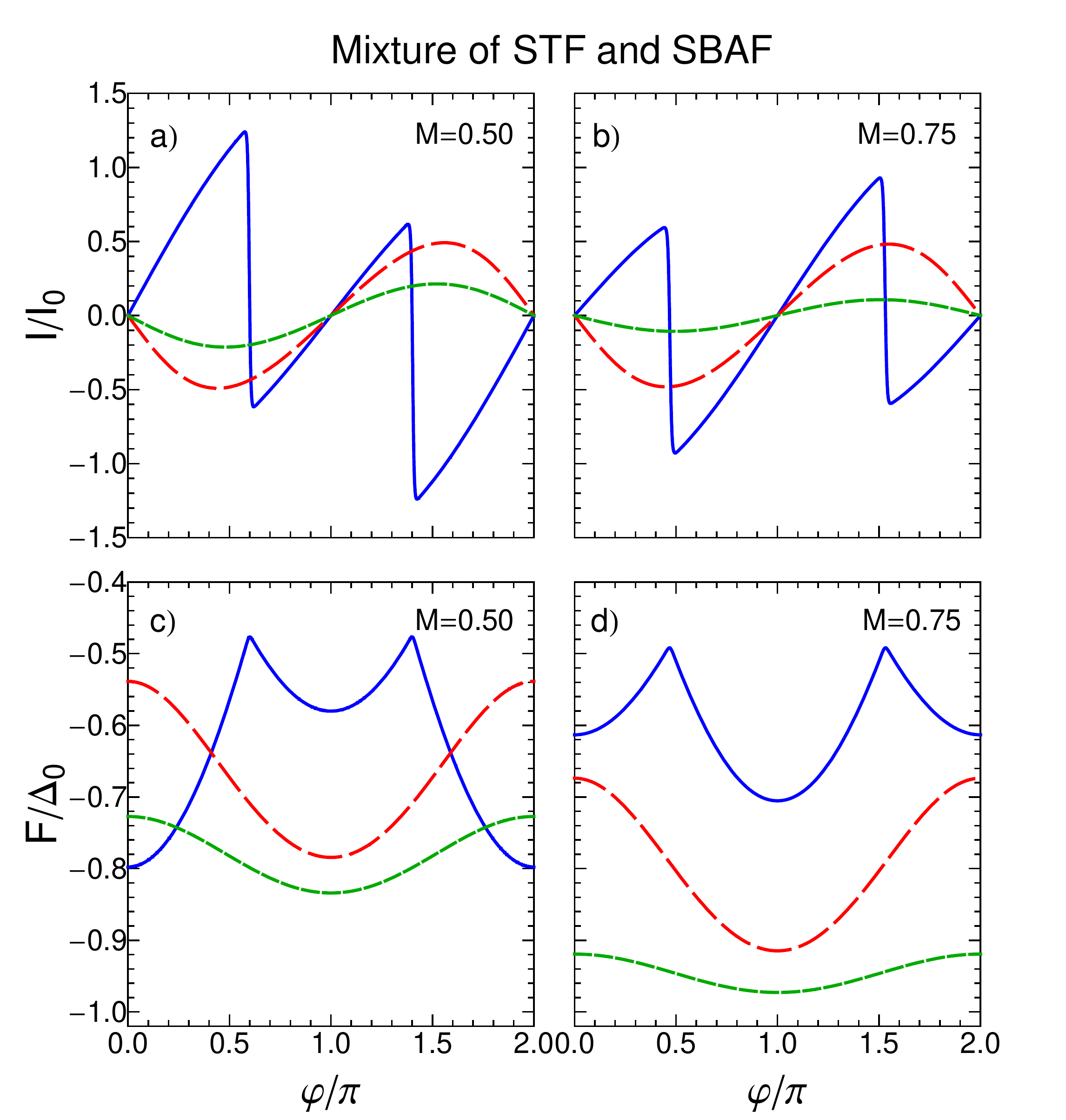}
\caption{(Color online). Josephson current and free energy for a
ferromagnet which is a degree $W=0.5$ mixture of STF and SBAF. All
parameters are fixed at the same values of
Fig.~\ref{fig:currentenergy}.} \label{fig:mixed}
\end{center}
\end{figure}

We now consider the case of a mixed F with both exchange energy
and bandwidth asymmetry. We show that the previously
discussed features distinguishing SBAF from STF as seen in
Fig.~\ref{fig:currentenergy}, are also present when considering a mixed F.
Indeed the Josephson current and the free energy for mixed F look
very similar to the case of a pure SBAF, when a suitable
contribution of bandwidth asymmetry to total polarization is
reached. Fig.~\ref{fig:mixed} shows the case of a mixture of SBAF
and STF with $W=0.5$ and all other parameters fixed as in
Fig.~\ref{fig:currentenergy}. In the mixed case both exchange and
masses ratio have non trivial values. The values assumed by these
parameters for the particular choice $W=0.5$ are reported in
Table~\ref{table}. As previously discussed for $M=0.75$, STF is in
a $0-$phase while SBAF is in a $\pi-$phase. It is clear from
panels $b)$ and $d)$ of Fig.~\ref{fig:mixed} that the particular
mixed case considered with $W=0.5$ is in a $\pi-$phase just like the pure
SBAF case (corresponding to $W=1$). Depending on the junctions parameters, a mixed F can behave as a pure SBAF even at smaller bandwidth asymmetry contributions. For the
particular case considered here, the mixed F behaves as a pure SBAF for
$W \gtrsim 0.3$. By comparing SBAF and STF at $M=0.5$, we have pointed
out that STF and is in a $\pi-$phase while SBAF is in a $0-$phase
for $Z=0$, while for smaller interface transparency SBAF switch to
the same phase of STF. Panels $a)$ and $c)$ of
Fig.~\ref{fig:mixed} show that even in this case a partial
bandwidth asymmetry in F is enough to observe features of a pure
SBAF. For the polarization considered ($M=0.5$), a mixed F behaves as a pure SBAF for $W \gtrsim 0.4$. These results are a clear signature
of the fact that the bandwidth asymmetry can induce unusual
features also when it is not the only mechanism producing the spin
polarization. We stress that this is an important consideration
since, as discussed later, in real ferromagnetic mid-layers both
exchange and bandwidth asymmetry mechanism may be present.

\begin{figure}
\begin{center}
\includegraphics[width=0.5\textwidth]{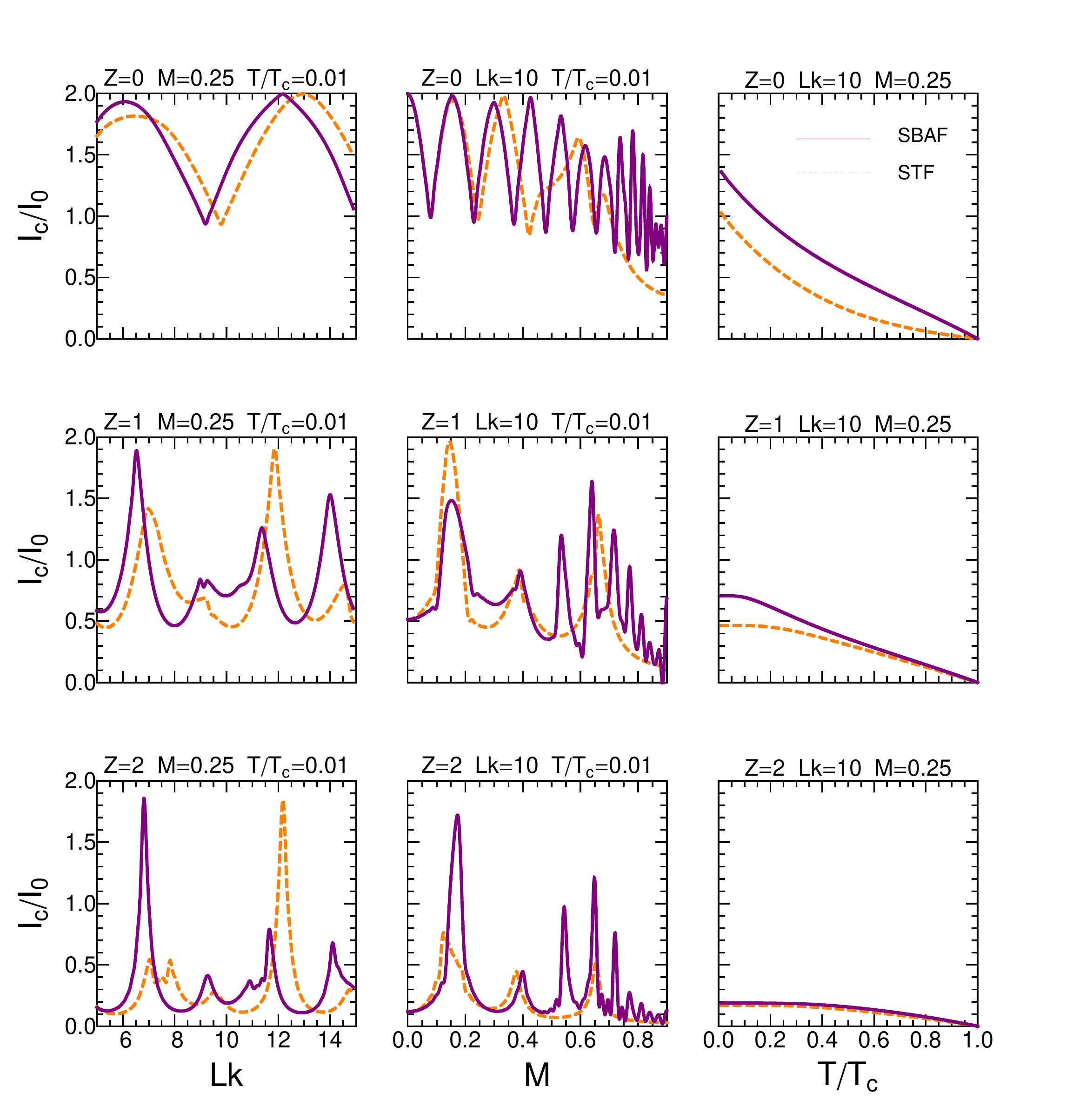}
\caption{(Color online). Critical Josephson current as a function
of $L$, $M$, and $T$ for SBAF (solid lines) and STF (dashed
lines).} \label{fig:criticalcurrent}
\end{center}
\end{figure}

Fig.~\ref{fig:criticalcurrent} shows the Josephson critical
current as a function of $L$, $M$, and $T$ for SBAF (solid lines)
and STF (dashed lines) for $Z=0,1,2$. When the $L$ dependence is
considered with a finite polarization, e.g.~$M=0.25$, the current
for both the STF and SBAF junction appears oscillatory for
transparent barriers and displays complicated patterns when $Z
\sim 1$. These patterns repeat periodically when $I_c$ is observed
on a larger scale and collapse to a modulated Dirac comb for $Z
\sim 10$. The maximum critical currents for the two junctions are
very close, but they are attained for different $L$ values and
appear to be weakly affected by the $Z$ value. The $M$ dependence
of the critical current shows a damped oscillatory behavior for
SBAF and STF for $Z=0$ and more complicated patterns for $Z=1,2$.
The main difference between the two junctions is that for high
polarization values, $M \gtrsim 0.6$, the critical current
decreases monotonically for STF, whereas for SBAF it oscillates
rapidly with increasing frequency. This is a signature of the fact
that SBAF still undergoes $0-\pi$ transitions, while STF has
settled in either a $0-$ or a $\pi-$phase. This behavior is
consistent with Eqs.~\eqref{eq:Istf} and~\eqref{eq:Isbaf} even if
they are strictly valid only in the tunneling limit. The plots of
the temperature dependence of the critical current in
Fig.~\ref{fig:criticalcurrent} show no particular features. The
critical currents in both STF and SBAF are decaying functions of
temperature, but the current in the latter junction is larger.
This is not a general feature but depends on the $M$ and $L$
values chosen. In general one can have both
$I_c^{\text{SBAF}}>I_c^{\text{STF}}$ and
$I_c^{\text{SBAF}}<I_c^{\text{STF}}$ as clearly shown by the plots
of critical current as a function of these parameters in
Fig.~\ref{fig:criticalcurrent}. Moreover, a temperature driven
$0-\pi$ transition can be seen both for STF and SBAF by carefully
tuning $M$ and $L$,~\cite{Petkovic} e.g.~$M=0.675$, $Lk=11.4$ for
SBAF and $M=0.935$, $Lk=11.4$ for STF. Even a slight change in one
of the two parameters suffices to destroy a temperature driven
transition, and the current simply decays as in
Fig.~\ref{fig:criticalcurrent}.

\begin{figure}
\begin{center}
\includegraphics[width=0.5\textwidth]{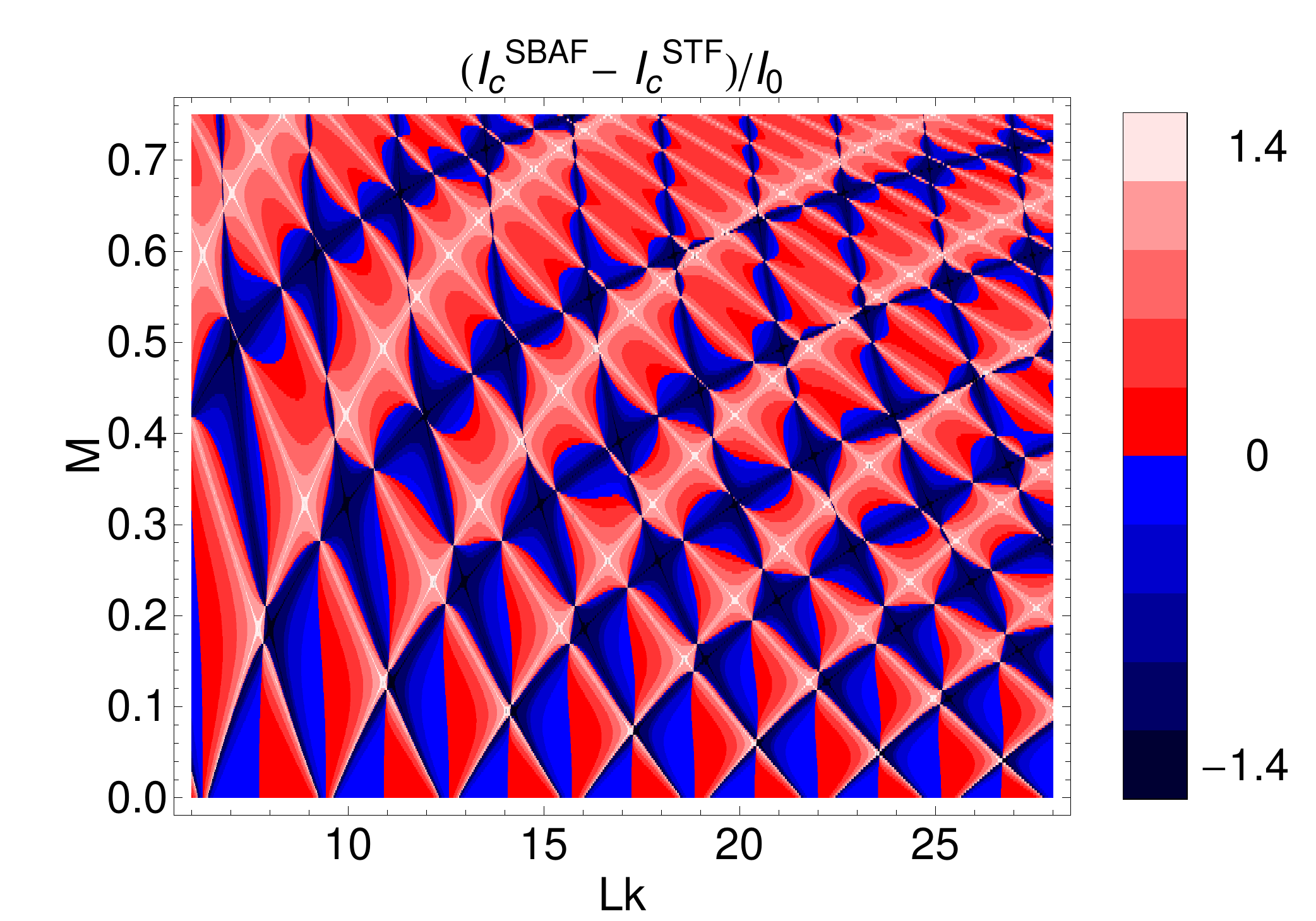}
\caption{(Color online). Density plot of $ \Delta I =(I_c^{SBAF}-
I_c^{STF}) / I_0 $ in the parameter range $0<M<0.75 \otimes
6<Lk<28$, and for $T=0$ and $Z = 10$. Red-light regions are for
$\Delta I >0$. The maximum value reached is $\Delta I \simeq 1.4$
Blue-dark regions are for $\Delta I <0$. The minimum value reached
is $\Delta I \simeq - 1.4$. At the borders of red and blue regions
the critical currents are equal and $\Delta I =0$.}
\label{fig:sbafgtstf}
\end{center}
\end{figure}

The analysis of the free energy and critical current has shown
that for a given set of junction parameters such as degree of
polarization, width, and insulating barrier strength, the
Josephson effect in SBAF and STF can be different. These two kinds
of ferromagnets can support different critical current magnitudes,
can induce different ground state phases across the junction, and
can undergo a different number of $0-\pi$ transitions driven by
variations in junction width and polarization. In order to study
these points in more detail, it is convenient to fix a range for
these parameters and then look at the behavior of STF and SBAF in
this entire range. We employ this strategy because the observables
are rapidly varying functions of $L$ and $M$. We here choose the
parameter range $ 6 \leq Lk \leq 28 \otimes 0 \leq M\leq 0.75 $ to
be consistent with both our approximations and with typical
experimental situations. First, we look at which of the
ferromagnetic mechanisms that gives rise to the larger critical
current, by considering the quantity $\Delta I \equiv
(I_c^{\text{SBAF}}- I_c^{\text{STF}})/ I_0$. This quantity is
plotted for the chosen range of parameters for $T=0$ and $Z=10$ in
Fig.~\ref{fig:sbafgtstf} where red-light regions are for $\Delta I
>0$. The maximum value reached is $\Delta I \simeq 1.4$. Blue-dark
regions are for $\Delta I <0$, where the minimum value reached is
$\Delta I \simeq - 1.4$. At the borders of red and blue regions,
the critical currents are equal and $\Delta I =0$. The complicated
pattern of $\Delta I$ shows that even a slight change in mid-layer
width or polarization can reverse the order of critical current
magnitudes in SBAF and STF meaning that $\Delta I$ has a strong
local character. In the low polarization regime both
$\Delta I <0$ and $\Delta I >0$ are realized with the same
frequency, while in most of the intermediate/high polarization regime $\Delta I>0$. This indicates that a SBAF is likely to support a larger
critical current than a STF when considering intermediate/strong ferromagnets. This property is qualitatively independent of
width, temperature, and interface transparency.

\begin{figure}
\begin{center}
\includegraphics[width=0.5\textwidth]{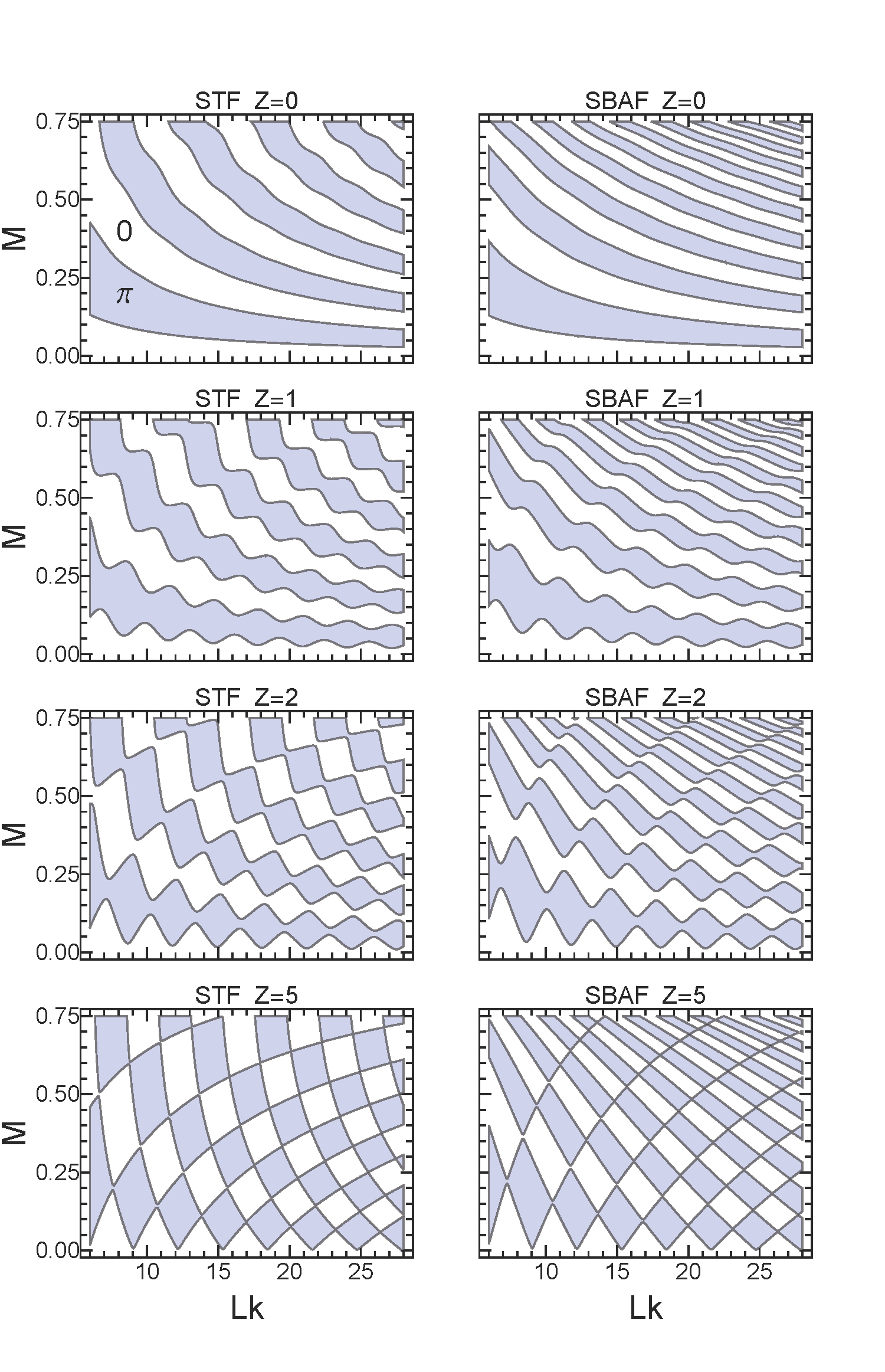}
\caption{(Color online). $M-L$ phase diagram for STF (left panels)
and SBAF (right panels) for $T=0.01 T_c$ and different $Z$ values.
White and colored regions correspond to $0-$ and $\pi-$phases,
respectively.} \label{fig:phasespace}
\end{center}
\end{figure}

We now analyze the $0-\pi$ transitions in the same range by
plotting phase diagrams for STF and SBAF at $T=0.01 T_c$ and
different values of $Z$ (see Fig.~\ref{fig:phasespace}). For the
moment, we choose to leave aside temperature dependence, since it
is not the main mechanism driving the $0-\pi$
transitions.~\cite{Petkovic}

From Fig.~\ref{fig:phasespace} it is clear that a larger number of
$M$ and $L$ driven $0-\pi$ transitions is generally expected for
SBAF than for STF. In particular, this is always the case for
polarization driven transitions at fixed width $L$,  regardless of
the value of $Z$. For $L$-driven transitions at fixed
polarization, this is the case only when $M \gtrsim 0.6$. Another
point to notice is that for both STF and SBAF the number of
possible transitions is larger for higher $Z$ values. At this
point the origin of the complicated patterns seen in
Fig.~\ref{fig:criticalcurrent} for $M$ and $L$ dependence of $I_c$
when $Z \neq 0$ is clear since $0-\pi$ region boundaries become
wavy for finite interface barrier strengths such that changing
only $M$ or $L$ leaving the other fixed, boundaries can be crossed
at a non-uniform frequency. We underline that the phase of the
junctions manifests a strong local character and can be switched
by slightly altering the width and/or the polarization.

\begin{figure}
\begin{center}
\includegraphics[width=0.5\textwidth]{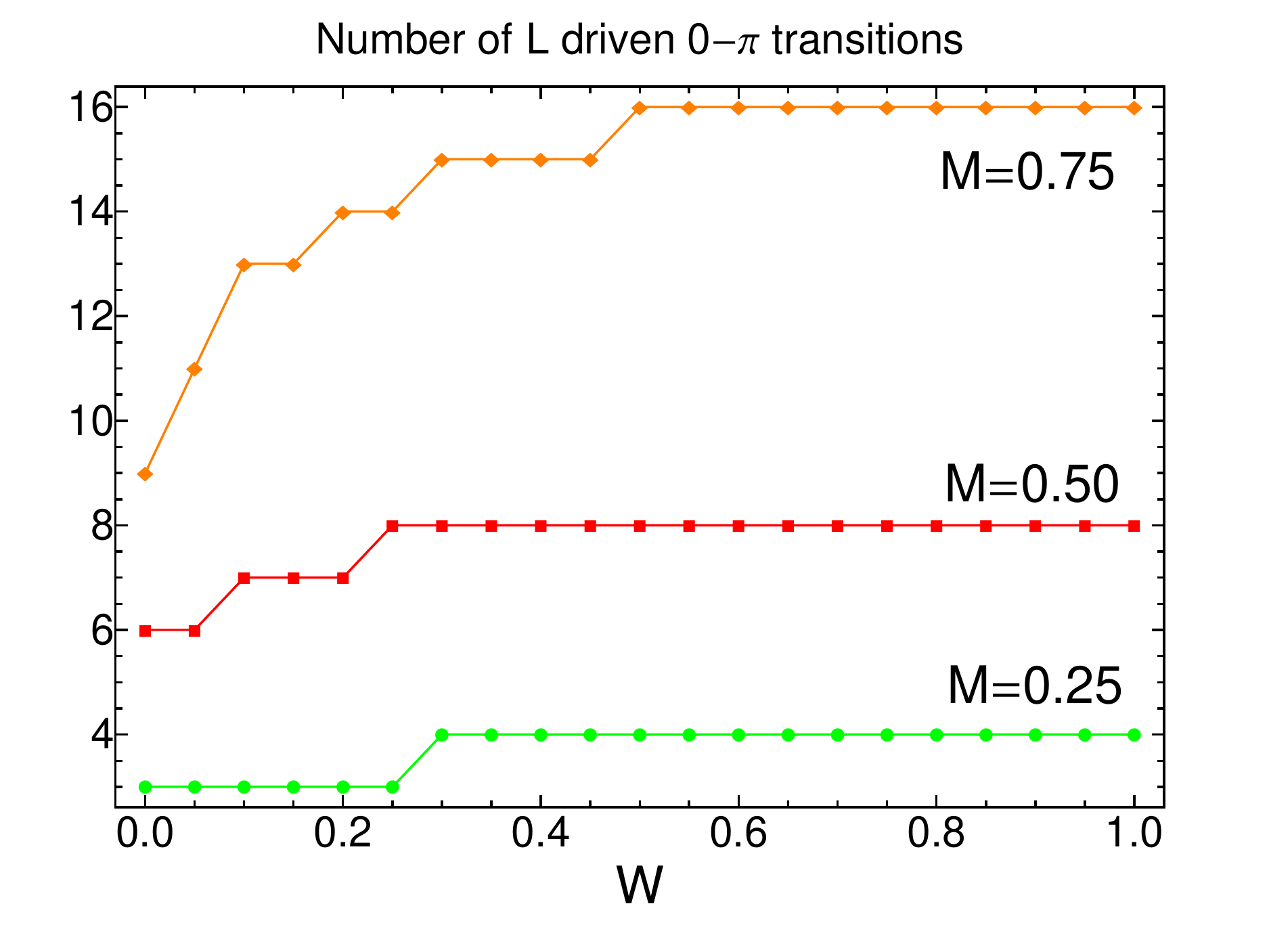}
\caption{(Color on line). Number of $L$ driven $0-\pi$ transitions
in the range $6<Lk<28$ as a function of the relative contribution
of SBAF to total polarization. $T/T_c=0.01$ and $Z=0$ are fixed.}
\label{fig:transnumb}
\end{center}
\end{figure}

Fig.~\ref{fig:phasespace} shows that SBAF undergoes more
transitions than STF. It is interesting to analyze the number of
transitions for a general F which can be a mixture of STF and
SBAF. Fig.~\ref{fig:transnumb} shows the number of $L$ driven
$0-\pi$ transitions at fixed polarization for different mixing
degree $W$ ranging from STF to SBAF. It is shown how the bandwidth
asymmetry tends to increase the number of transitions and that
this effect is more prominent in the high polarization regime
where the number of transitions is almost doubled. One important
point to note is that the same number of transitions observed in a
SBAF ($W=1$) can be observed in a mixed F whenever $W$ is large
enough, e.g.~$W=0.5$ for $M=0.75$. This is an important point when
relating our results with real materials because in experiments,
one usually encounters the mixed F case. Even if $T/T_c=0.01$ and 
$Z=0$ are fixed in Fig.~\ref{fig:phasespace}, we
underline that the general trend of bandwidth asymmetry increasing
the number of transitions does not depend on their particular
values.


\section{CONCLUSIONS}

In conclusion, we have analyzed the Josephson effect in short,
ballistic single channel S/F/S junctions taking into account the
possibility for ferromagnetism to be driven by a mass
renormalization of carriers with opposite spin, i.e. a spin
bandwidth asymmetry. We have compared a junction with this
unconventional kinetically driven ferromagnetism in the F layer
with one with the usual Stoner mechanism considering also their
interplay. Analyzing Andreev levels, free energy and currents, we
have shown that the Josephson effect in the two junctions shows
different features especially for intermediate/high polarization
values. In particular, we have shown that the junction with total
or partial spin bandwidth asymmetry in the F layer undergoes a
larger number of $0-\pi$ transitions driven by variations in
junction width and polarization. By examining free energy and
phase diagrams, we have pointed out how junctions with different
magnetic mechanisms in the F layer can be in different phases even
if all junction parameters have the same values. We have remarked
that this is a common (rare) situation in the intermediate/high
(low) polarization regime. By analyzing Josephson critical current
we have shown that bandwidth asymmetry can both enhance and
decrease its value, the former situation being more common for
strong ferromagnets. Our findings are relevant for many
interesting magnetic materials which cannot be framed
exclusively within a Stoner scenario. As relevant examples we cite
the half-metal ferromagnets defined by the property of having almost
$100 \%$ transport spin polarization.~\cite{Mazin99} Materials
belonging to this class usually have spin polarization $M$ in the
range where bandwidth asymmetry is clearly distinguishable from
Stoner exchange, e.g.~ CrO$_2$,~\cite{Chioncel2007}
La$_{0.7}$Sr$_{0.3}$MnO$_3$,~\cite{Nadgorny2001}
Fe$_3$O$_4$,~\cite{Dedkov2002} and EuO,~\cite{Steen2002} among
others. Their high polarization values cannot be ascribed only to
Stoner exchange, so that when trying to theoretically model them
the inclusion of bandwidth asymmetry is a fundamental ingredient.

We would like to emphasize that our findings may turn out to be 
useful in ascertaining if a given ferromagnetic material has
spin bandwidth asymmetry. Furthermore, since a bandwidth
asymmetric F layer can provide more frequent transitions and is
likely to support a larger critical current for a given set of
parameters than a Stoner ferromagnet, it may be more suitable for
electronics or spintronics applications and devices based on the
Josephson effect and relying on $0-\pi$ transitions. Moreover,
considering experimental and sample production limitations, the
choice of magnetization mechanism may work as an extra degree of
freedom which by itself may determine the ground state phase
difference across the junction or the magnitude of Josephson
critical current.

\acknowledgments
A. S. was supported by the Norwegian Research Council under Grant No. 167498/V30 (STORFORSK).

\appendix
\section{Explicit form of Andreev levels}
\label{app1}
The explicit form of $A,B,C,D$ in
Eq.~\eqref{eq:levels} is
\begin{align}
    A &= 2(a-b) \;, \\
    B &= 4c         \;, \\
    C &= 2(a+b)-d   \;, \\
    D &= 2d         \;,
\end{align}

where

\be
    a = -\left(\lambda_{\sigma}^2 + \lambda_{\bar{\sigma}}^2\right)
    - 2\frac{\lambda_{\sigma}}{\tan{(kL\Lambda_{\sigma})}}\frac{\lambda_{\bar{\sigma}}}
    {\tan{(kL\Lambda_{\bar{\sigma}})}}  ,
\ee
\begin{multline}
    b = \left(1+Z^2\right)^2
    -Z^2\left(\lambda_{\sigma}^2+\lambda_{\bar{\sigma}}^2\right) \\
    +\left(4Z^2 + 2\right)\frac{\lambda_{\sigma}}{\tan{(kL\Lambda_{\sigma})}}
    \frac{\lambda_{\bar{\sigma}}}{\tan{(kL\Lambda_{\bar{\sigma}})}} +
    \lambda_{\sigma}^2\lambda_{\bar{\sigma}}^2 \\
    +2Z\left(1+Z^2\right)\left(\frac{\lambda_{\sigma}}{\tan{(kL\Lambda_{\sigma})}} +
    \frac{\lambda_{\bar{\sigma}}}{\tan{(kL\Lambda_{\bar{\sigma}})}}\right) \\-
    2Z\left(\lambda_{\bar{\sigma}}^2\frac{\lambda_{\sigma}}{\tan{(kL\Lambda_{\sigma})}}
     + \lambda_{\sigma}^2\frac{\lambda_{\bar{\sigma}}}{\tan{(kL\Lambda_{\bar{\sigma}})}} \right) ,
\end{multline}

\begin{multline}
    c = \left(1+Z^2\right)\left(\frac{\lambda_{\sigma}}
    {\tan{(kL\Lambda_{\sigma})}} - \frac{\lambda_{\bar{\sigma}}}
    {\tan{(kL\Lambda_{\bar{\sigma}})}}\right) \\+
    \lambda_{\bar{\sigma}}^2(1+Z)\frac{\lambda_{\sigma}}{\tan{(kL\Lambda_{\sigma})}}
    - \lambda_{\sigma}^2(1+Z)\frac{\lambda_{\bar{\sigma}}}{\tan{(kL\Lambda_{\bar{\sigma}})}},
\end{multline}

and \be
    d = 8\frac{\lambda_{\sigma}}{\sin{(kL\Lambda_{\sigma})}}
    \frac{\lambda_{\bar{\sigma}}}{\sin{(kL\Lambda_{\bar{\sigma}})}}
    ,
\ee

where
\begin{align}
    \lambda_{\sigma} &= \sqrt{\frac{m}{m_{\sigma}}\left(1+\rho_{\sigma}\frac{U}{E_F}\right)} , \\
    \Lambda_{\sigma} &= \sqrt{\frac{m_{\sigma}}{m}\left(1+\rho_{\sigma}\frac{U}{E_F}\right)}
    .
\end{align}

\section{Determination of magnetic parameters for mixed F}
\label{app2}

Since the polarization is not a separable function with respect to
the exchange interaction and mass mismatch, one cannot immediately
obtain the values assumed by these microscopic parameters when F
is a mixture of STF and SBAF. In this appendix we illustrate how
we perform this calculation for a given $M$ and $W$, where the
mixing parameter $W$ quantifies the relative weight of the two
contributions to $M$. We notice that STF corresponds to $W=0$ and
SBAF to $W=1$.

Let us consider the two dimensional $(m_\uparrow/m_\downarrow,U/E_F)$
space parameter over which $M$ is defined (see e. g. Fig 1 of
Ref.~\cite{Annunziata}). For a given value of the polarization $M$,
and from Eq.~\eqref{eq:U}, STF is represented by the point
$\left(\frac{2M}{(1+M^2)},0 \right)$ and, from Eq.~\eqref{eq:y},
SBAF is represented by the point
$\left(1,\left(\frac{M+1}{M-1}\right) ^2 \right)$. The distance
between these two points along a constant magnetization path
$l(M)$ can be written as

\begin{eqnarray*}
l(M) & = & \frac{e^{i \pi/4}}{2\sqrt{2}} \left[ B \left(-\frac{1}{4} \left( \frac{M+1}{M-1} \right)^4 ;-1/4,3/2 \right)\right. \nonumber \\
 & & {}
 \left.-B\left(-4\left(\frac{M^2+1}{M^2-1}\right)^4;-1/4,3/2\right)\right],
\end{eqnarray*}
where $B$ is the incomplete Euler function \be \nonumber
B(z;a,b)=\int_0^z t^{a-1}(1-t)^{b-1} dt . \ee Then the coordinates
in parameter space associated with a mixture defined by the actual
$W$ can be evaluated by imposing that the path between the STF and
the sought points is a fraction of the total length equals to $W\,
l(M)$. Consequently, the mass mismatch value for a degree $W$
mixed F at polarization $M$ can be estimated by numerically
solving the equation

\begin{eqnarray*}
 & &\frac{e^{i \pi/4}}{2\sqrt{2}} \left[ B \left(f\left(\frac{m_\uparrow}{m_\downarrow},M \right);-1/4,3/2 \right)\right. \nonumber \\
 & & {}\left.-B\left(-4\left(\frac{M^2+1}{M^2-1}\right)^4;-1/4,3/2\right)\right]= W \, l(M),
\end{eqnarray*}
where

\be \nonumber f\left(\frac{m_\uparrow}{m_\downarrow},M\right) =
-\frac{1}{4} \left(
\frac{(M+1)^2+\frac{m_\uparrow}{m_\downarrow}(M-1)^2}{M^2-1}
\right)^4. \ee Once the sought value of mass mismatch is known,
the corresponding value of exchange interaction can be obtained by
inversion of  Eq.~\eqref{eq:mag}.


\begin{thebibliography}{999}



\bibitem{Fulde} P. Fulde and R. A. Ferrell,
Phys. Rev. {\bf 135}, 550 (1964).

\bibitem{Larkin} A. I. Larkin and Y. N. Ovchinnikov, Zh. Eksp. Teor. Fiz. 47, 1136
(1964) [Sov. Phys. JETP 20, 762 (1965)].

\bibitem{Maska} M. M. Ma\'s ka, M. Mierzejewski, J. Kaczmarczyk, and J. Spalek,
Phys. Rev. B {\bf 82}, 054509 (2010).

\bibitem{Romano2010} A. Romano, M. Cuoco, C. Noce, P. Gentile, and G. Annunziata,
Phys. Rev. B {\bf 81}, 064513 (2010).

\bibitem{Soulen98} R. J. Soulen Jr., J. M. Byers, M. S. Osofsky, B. Nadgorny,
T. Ambrose, S. F. Cheng, P. R. Broussard, C. T. Tanaka, J. Nowak,
J. S. Moodera,    A. Barry, and J. M. D. Coey, Science {\bf 282},
85 (1998); S. K. Upadhyay, A. Palanisami, R. N. Louie, and R. A.
Buhrman, Phys. Rev. Lett. {\bf 81}, 3247 (1998).

\bibitem{Tedrow94} R. Meservey and P.M. Tedrow, Phys. Rep. {\bf
238}, 173 (1994).

\bibitem{spintronicsreview} I. \v{Z}uti\'c, J. Fabian, and S.D. Sarma,
Rev. Mod. Phys. {\bf 76}, 323 (2704).

\bibitem{Andreev64} A. F. Andreev, Zh. Eksp. Teor. Fiz. {\bf 46}, 1823 (1964) [Sov. Phys. JETP {\bf 19}, 1228 (1964)].

\bibitem{Beenakker92} C. W. J. Beenakker,
Phys. Rev. B {\bf 46}, 12841 (1992).

\bibitem{deJong} M. J. M de Jong and C. W. J. Beenakker, Phys. Rev. Lett. {\bf 74}, 1657 (1995).

\bibitem{linder_prb_07} S. Kashiwaya, Y. Tanaka, N. Yoshida, and M. R. Beasley, Phys. Rev. B \textbf{60}, 3572 (1999);
Igor Zutic and Oriol T. Valls, Phys. Rev. B \textbf{60}, 6320 (1999);
J. Linder and A. Sudb{\o}, Phys. Rev. B \textbf{75}, 134509 (2007).

\bibitem{Bulaevskii} L. N. Bulaevskii, V. V. Kuzii, and A. A. Sobyanin, JETP Lett. {\bf
25}, 290 (1977).

\bibitem{Ryazanov} V. V. Ryazanov, V. A. Oboznov, A. Yu Rusanov, A. V. Veretennikov,
A. A. Golubov, and J. Aarts, Phys. Rev. Lett. {\bf 86}, 2427
(2001).

\bibitem{Kontos} T. Kontos, M. Aprili, J. Lesueur, and X. Grison,
Phys. Rev. Lett. {\bf 86}, 304 (2001).

\bibitem{Yamashita} T. Yamashita, K. Tanikawa, S. Takahashi, and
S. Maekawa, Phys. Rev. Lett. {\bf 95}, 097001 (2005).

\bibitem{Chtchelkatchev} M. Chtchelkatchev, W. Belzig, Yu. V. Nazarov, and C. Bruder, JETP Lett. {\bf 74}, 323 (2001).

\bibitem{Radovic} Z. Radovic, N. Lazarides, and N. Flytzanis,
Phys. Rev. B {\bf 68},014501 (2003).

\bibitem{Cayssol} J. Cayssol, and G. Montambaux,
Phys. Rev. B {\bf 70}, 224520 (2004).


\bibitem{Petkovic} I. Petkovic, N. M. Chtchelkatchev, and Z. Radovic,
Phys. Rev. B {\bf 73}, 184510 (2006).

\bibitem{linder_prl_08} J. Linder, T. Yokoyama, D. Huertas-Hernando, and A. Sudb{\o}, Phys. Rev. Lett. \textbf{100}, 187004 (2008).


\bibitem{Millis88} A. Millis, D. Rainer, and J. A. Sauls, Phys. Rev. B {\bf 38},
4504 (1988).

\bibitem{Fogel2000} M. Fogelstr\"om, Phys. Rev. B {\bf 62}, 11 812 (2000).


\bibitem{Barash2002} Yu. S. Barash and I. V. Bobkova, Phys. Rev. B {\bf 65}, 144502
(2002).

\bibitem{josreview} K. K. Likharev, Rev. Mod. Phys. {\bf 51}, 101 (1979).

\bibitem{CPRreview} A. A. Golubov, M.Yu. Kupriyanov, and E. Il'ichev, Rev. Mod.
Phys. {\bf 76}, 411 (2004).


\bibitem{Abinitio} R. Strack, and D. Vollhardt, Phys. Rev. Lett. {\bf 72}, 3425
(1994); M. Kollar, R. Strack, and D. Vollhardt, Phys. Rev. B {\bf
53}, 9225 (1996); D. Vollhardt, N. Bl\"umer, K. Held, J. Schlipf,
and M. Ulmke, Z. Phys. B {\bf 103}, 283 (1997); J. Wahle, N.
Bl\"umer, J. Schlipf, K. Held, and D. Vollhardt, Phys. Rev. B {\bf
58}, 12749 (1998).

\bibitem{Hirsch} J.E. Hirsch, Phys. Rev. B {\bf 40}, 2354 (1989);
J.E. Hirsch, {\it ibid.} {\bf 40}, 9061 (1989); J.E. Hirsch, {\it
ibid.} {\bf 43}, 705 (1991); J.E. Hirsch, {\it ibid.} {\bf 62},
14131 (2000); J.E. Hirsch, Physica C {\bf 341-348}, 211 (2000).


\bibitem{Hirsch99} J.E. Hirsch, Phys. Rev. B, {\bf 59},
6256 (1999).


\bibitem{Okimoto} Y. Okimoto, T. Katsufuji, T. Ishikawa, A. Urushibara, T. Arima,
and Y. Tokura, Phys. Rev. Lett. {\bf 75}, 109 (1995); Y. Okimoto, T. Katsufuji,
T. Ishikawa, T. Arima, and Y. Tokura, Phys. Rev. B {\bf 55}, 4206 (1997).

\bibitem{hexab} L. Degiorgi, E. Felder, H.R. Ott, J.L. Sarrao, and Z. Fisk,
Phys. Rev. Lett. {\bf 79}, 5134 (1997); S. Broderick, B. Ruzicka,
L. Degiorgi, H.R. Ott, J. L. Sarrao, and Z. Fisk, Phys. Rev. B {\bf
65}, 121102 (2002).

\bibitem{Singley} E.J. Singley, K.S. Burch, R. Kawakami, J. Stephens, D.D.
Awschalom, and D.N. Basov, Phys. Rev. B {\bf 68}, 165204 (2003);
E.J. Singley, R. Kawakami, D.D. Awschalom, and D.N. Basov, Phys.
Rev. Lett. {\bf 89}, 097203 (2002).


\bibitem{Ying} Z.-J. Ying,
M. Cuoco, C. Noce, and H.-Q. Zhou, Phys. Rev. B {\bf 78}, 104523
(2008); M. Cuoco, P. Gentile, and C. Noce, Phys. Rev. Lett. {\bf
91}, 197003 (2003).

\bibitem{Cuoco} M. Cuoco, A. Romano, C. Noce,
and P. Gentile, Phys. Rev. B {\bf 78}, 054503 (2008).

\bibitem{Annunziata} G. Annunziata, M. Cuoco, C. Noce, A. Romano,
and P. Gentile, Phys. Rev. B {\bf 80}, 012503 (2009).

\bibitem{Annunziata2010} G. Annunziata, M. Cuoco, P. Gentile, A. Romano,
and C. Noce, Phys. Rev. B, {\bf 83}, 094507 (2011). 



\bibitem{Kac2009} J. Kaczmarczyk and J. Spa\l ek, Phys. Rev. B {\bf 79}, 214519 (2009).

\bibitem{Maska2010} M. M. Ma\'{s}ka, M. Mierzejewski, J. Kaczmarczyk, and J. Spa\l ek, Phys. Rev. B {\bf 82}, 054509 (2010).



\bibitem{BdG} P. G. de Gennes, in {\it Superconductivity of
Metals and Alloys}, (W.A. Benjamin, Inc. New York, 1966).

\bibitem{AnnunziataSUST} G. Annunziata, M. Cuoco, P. Gentile, A. Romano, and C. Noce, Supercond. Sci. Technol. {\bf 24}, 024021  (2011).

\bibitem{Beenakker912} C. W. J. Beenakker, Phys. Rev. Lett. {\bf 67}, 3836
(1991).

\bibitem{Kulik} I. O. Kulik, Zh. Eksp. Teor. Fiz. {\bf 57}, 1745 (1969) [Sov. Phys. JETP {\bf 30}, 944 (1970)].

\bibitem{Beenakker91} C. W. J. Beenakker and H. van Houten, Phys. Rev. Lett. {\bf 66}, 3056
(1991).



\bibitem{kwon} H. J. Kwon, V. M. Yakovenko, and K. Sengupta, Low Temp. Phys. {\bf 30}, 613 (2004).

\bibitem{baselmans} J. J. A. Baselmans, T. T. Heikkil{\"a}, B. J. van Wees and T. M. Klapwijk,
Phys. Rev. Lett. {\bf 89}, 207002, (2002).

\bibitem{goldobin} E. Goldobin, D. Koelle, R. Kleiner, and A. Buzdin, Phys. Rev. B {\bf 76}, 224523 (2007).





\bibitem{Mazin99} I.I. Mazin, Phys. Rev. Lett. {\bf 83}, 1427 (1999).

\bibitem{Chioncel2007} L. Chioncel, H. Allmaier, E. Arrigoni, A. Yamasaki, M. Daghofer,
M. Katsnelson, and A. Lichtenstein, Phys. Rev. B {\bf 75}, 140406
(2007).

\bibitem{Nadgorny2001} B. Nadgorny, I. I. Mazin, M. Osofsky, R. J. Soulen, Jr., P.
Broussard, R. M. Stroud, D. J. Singh, V. G. Harris, A. Arsenov,
and Y. Mukovskii, Phys. Rev. B {\bf 63}, 184433 (2001).

\bibitem{Dedkov2002} Y. S. Dedkov, U. R\"{u}diger, and G. G\"{u}ntherodt, Phys. Rev. B {\bf 65},
064417 (2002).

\bibitem{Steen2002} P. G. Steeneken, L. H. Tjeng, I. Elfimov, G. A. Sawatzky, G.
Ghiringhelli, N. B. Brookes, and D.-J. Huang, Phys. Rev. Lett.
{\bf 88}, 047201 (2002).








\end{thebibliography}
\end{document}